\documentclass{elsart}
\usepackage{amsmath,amssymb}
\usepackage{graphics}
\usepackage{graphicx}
\usepackage{epsfig}

\def\build#1_#2^#3{\mathrel{\mathop{\kern 0pt#1}\limits_{#2}^{#3}}}

\begin{document}

\begin{frontmatter}



\title{Unified Multifractal Description of Velocity Increments Statistics in Turbulence: Intermittency and Skewness}
\author[JHU,ENS]{L. Chevillard\corauthref{cor}},
\corauth[cor]{Corresponding author.} \ead{chevillard@jhu.edu}
\author[ENS]{B. Castaing}
\author[ENS]{, E. L\'ev\^eque}
\author[ENS]{, A. Arneodo}

\address[JHU]{Department of Mechanical Engineering, the Johns Hopkins University, \\3400 N. Charles Street, Baltimore, MD 21218, USA}
\address[ENS]{Laboratoire de Physique, \textsc{cnrs}, \'Ecole Normale
Sup\'erieure de Lyon, 46 all\'ee d'Italie, 69364 lyon 7, France}


\author{}

\address{}

\begin{abstract}
The phenomenology of velocity statistics in turbulent flows, up to
now, relates to different models dealing with either signed or
unsigned longitudinal velocity increments, with either inertial or
dissipative fluctuations. In this paper, we are concerned with the
complete probability density function (PDF) of signed longitudinal
increments at all scales. First, we focus on the symmetric part of
the PDFs, taking into account the observed departure from scale
invariance induced by dissipation effects. The analysis is then
extended to the asymmetric part of the PDFs, with the specific
goal to predict the skewness of the velocity derivatives. It opens
the route to the complete description of all measurable
quantities, for any Reynolds number, and various experimental
conditions. This description is based on a single universal
parameter function $\mathcal D(h)$ and a universal constant
$\mathcal R^*$.

\end{abstract}

\end{frontmatter}

\section{Introduction}

In the field of  turbulence, a significant effort has been devoted
to the analysis of the scaling behavior of structure functions
$\langle (\delta_\ell u)^q\rangle $, where  $\delta_\ell u$ is the
longitudinal velocity increment between two points separated by a
variable distance $\ell$ \cite{Fri95}.  However, a better strategy
may be to concentrate on the probability density functions (PDFs)
of  $\delta_\ell u$, rather than on a set of moments
\cite{KaiSre}. Accordingly, this work deals with the description
of  the PDFs of $\delta_\ell u$, where the scale $\ell$ spans the
entire range of excited scales of motion (from the integral far
down to the dissipative scales). Experimental and numerical
observations have provided the evidence that the PDFs of
$\delta_\ell u$ are increasingly stretched as $\ell$ decreases,
while they are almost Gaussian at the large scales where the
turbulence is stired \cite{Fri95}. This feature is known as
intermittency. Moreover,  Chevillard \emph{et al.} \cite{CheCas03}
have recently argued that this stretching is largely enhanced in
the near-dissipation range, leading to extremely high fluctuation
level for the velocity gradients. Another essential feature lies
in the significant asymmetry, or skewness, of the PDFs. This
skewness to be non-zero is heuristically connected to the vortex
folding and stretching (irreversible) process, which drains energy
from the large to the small scales, and hence, plays a central
role in turbulence. From a theoretical viewpoint, a quantitative
description of the skewness is still missing. In this context, our
motivation is to present a synthetic description of the PDFs of
$\delta_\ell u$, which encompasses the combined effects of
intermittency and skewness. To do so, the PDF of $\delta_\ell u$
is split into a symmetric $\mathcal P^+_{\delta_\ell u}$ and an
asymmetric $\mathcal P^-_{\delta_\ell u}$ part. First, the focus
is on $\mathcal P^+_{\delta_\ell u}$, which also represents the
PDF of the magnitude of $\delta_\ell u$. We show that $\mathcal
P^+_{\delta_\ell u}$ is suitably described by a multifractal
picture of turbulence dynamics \cite{Fri95},  which incorporates
finite-Reynolds-number effects. The analysis is then extended to
$\mathcal P^-_{\delta_\ell u}$ with the specific goal to describe
the skewness phenomenon via a quantitative estimate of the
skewness factor $\langle \delta_\ell u^3 \rangle/\langle
\delta_\ell u^2 \rangle^{3/2}$ as a function of $\ell$.

\section{Statistics of longitudinal velocity increment magnitude: Modeling the symmetric part of the Probability Density Function}

From a general point of view, the PDF of the longitudinal velocity
increments $\mathcal P_{\delta_\ell u}$ can be split into a
symmetric $\mathcal P ^+_{\delta_\ell u}$ (i.e. even) function and
an asymmetric $\mathcal P ^-_{\delta_\ell u}$ (i.e. odd) function
in the following way:
\begin{equation}
\mathcal P_{\delta_\ell u}(\delta_\ell u) = \mathcal
P^+_{\delta_\ell u}(\delta_\ell u) + \mathcal P^-_{\delta_\ell
u}(\delta_\ell u)\mbox{ .}
\end{equation}
The PDF of the longitudinal velocity increment magnitude $\mathcal
P_{|\delta_\ell u|}(|\delta_\ell u|) = \mathcal P_{\delta_\ell
u}(|\delta_\ell u|) + \mathcal P_{\delta_\ell u}(-|\delta_\ell u|)
= 2 \mathcal P^+_{\delta_\ell u}(|\delta_\ell u|)$ shows that the
symmetric part of the PDF of $\delta_\ell u$ describes the
magnitude statistics. Notice that neither $\mathcal
P^+_{\delta_\ell u}$ nor $\mathcal P^-_{\delta_\ell u}$ can be
interpreted as a PDF of a random variable.

Let us first focus on the symmetric part of the PDFs of the
longitudinal velocity increments $\mathcal P^+_\ell (\delta_\ell
u)$. In the inertial range, the multifractal formalism
\cite{ParFri85}, which a priori pertains in the limit of infinite
Reynolds number, states that velocity is everywhere singular, the
longitudinal velocity increments at scale $\ell$ behaving locally
as $\ell^h$, where the H$\ddot{\mbox{o}}$lder exponent $h$ takes
value in a finite interval $[h_{\mbox{\small min}},h_{\mbox{\small
max}}]$. When the Reynolds number $\mathcal R_e = \sigma L/\nu$ is
finite ($L$ is the correlation length scale, $\sigma =
\sqrt{\langle (\delta_L u)^2\rangle} $ and $\nu$ the kinematic
viscosity), Paladin and Vulpiani \cite{PalVul87} have argued that
the  dissipative scale, that is supposed to separate the inertial
and the dissipation scaling ranges, is not unique in the presence
of intermittency and is likely to depend on $h$. Using these
arguments, Nelkin \cite{Nel90} predicted the moments of velocity
gradients, i.e. $\langle (\partial_x u)^q\rangle$. The
phenomenological consequences on the energy power spectrum were
studied by Frisch and Vergassola \cite{FriVer91} who proved the
existence of an intermediate dissipative range. Meneveau
\cite{Men96} further investigated the behavior of the structure
functions in that transitory range. Recently, Chevillard
\textit{et al.} \cite{CheCas03} revisited the behavior of
longitudinal velocity increments in the intermediate dissipative
range and showed, among other new predictions, that the width
$[\eta _-,\eta_+]$ of this range of scales behaves non trivially
with the Reynolds number, i.e. $\ln (\eta_+/\eta_-)\sim \sqrt{\ln
\mathcal R_e}$.

To provide a complete statistical description of longitudinal
velocity increments statistics, one needs to model the probability
law of the stochastic variable $|\delta_\ell u|$. As originally
proposed by Castaing \textit{et al.} \cite{CasGag90}, within the
propagator approach, velocity increments magnitude can be
considered as the product of two independent random variables,
$|\delta_\ell u| = \beta_\ell \times |\delta|$ (in law), where
$\delta$ is a zero-mean gaussian random variable of variance
$\sigma^2$ and $\beta_\ell$ a positive random variable (see
\cite{CheCas03} for details). In the inertial range, i.e. $\eta
(h) \ll \ell \ll L$, where $\eta(h)$ is the fluctuating
dissipative scale, $\beta_\ell(h) = (\ell/L)^h$ can be expressed
as a function of the singularity strength $h$ that fluctuates from
point to point according to the probability law $\mathcal
P_\ell(h) \sim (\ell/L)^{1-\mathcal D (h)}$. Note that the
exponent $h$ and the parameter function $\mathcal D(h)$ gain the
mathematical status of H$\ddot{\mbox{o}}$lder exponent and
singularity spectrum in the inviscid limit ($\mathcal R_e
\rightarrow +\infty$). The dissipative scale $\eta(h)$ fluctuates
according to : $\eta(h) = L(\mathcal R_e/\mathcal
R^*)^{-1/(h+1)}$, where the constant $\mathcal R^*$ is necessary
to be consistent with experimental and numerical data
\cite{CasGag93_1,CasGag93_2,CasGag93_3}. More precisely, in a
monofractal description of velocity fluctuations (Kolmogorov K41
theory \cite{Fri95}), one can show \cite{CheCas03} that the
Kolmogorov constant $c_K = \langle(\delta_\ell u)^2\rangle/\langle
\epsilon\rangle^{2/3}\ell^{2/3}=(\mathcal R^*/15)^{2/3}$. Actually
in this simplified monofractal framework, the average local
dissipation $\langle \epsilon\rangle = 15\nu \langle
(\partial_xu)^2\rangle = (15/\mathcal R^*)\times \sigma^3/L$  can
be related to the ratio $\mathcal R_e /(\mathcal R_\lambda)^2 =
4/\mathcal R^*$ (where $\mathcal R_\lambda$ is the Taylor based
Reynolds number). We will see in the following that the data are
compatible with the universal value $\mathcal R^*=52$ (in the
presence of intermittency).

For scales $\ell\le \eta(h)$, the velocity is smooth and Taylor's
development applies, i.e. $\delta_\ell u(x) = \ell\partial_xu(x)$.
In the multifractal description, using a simple continuity
argument with the inertial range behavior \cite{Nel90} yields
$\beta_\ell (h) = (\ell/L)(\eta(h)/L)^{h-1}$ and $\mathcal P_\ell
(h) \sim (\eta(h)/L)^{1-\mathcal D(h)}$. Then, we impose that the
function $\beta_\ell$ be continuous and differentiable at the
transition, following a strategy already used in a slightly
different form in Ref. \cite{Men96}, and which is inspired from an
elegant interpolation formula originally proposed by Batchelor
\cite{Bat51}, independently derived in a field theoretic approach
\cite{SirSmi}. In this framework, a single function $\beta_\ell(h,
\mathcal R_e/\mathcal R^*)$ covers the entire range of scale:
\begin{equation}\label{eq:betah}
\beta_\ell(h,\mathcal R_e/\mathcal R^*) =
\frac{\left(\frac{\ell}{L}\right)^h}{\left[
1+\left(\frac{\ell}{\eta(h)}\right)^{-2}\right]^{(1-h)/2}} \mbox{
,}
\end{equation}
and
\begin{equation}\label{eq:phmeneeul}
\mathcal P_\ell(h,\mathcal R_e/\mathcal R^*,\mathcal D)
=\frac{1}{\mathcal Z(\ell)}
\frac{\left(\frac{\ell}{L}\right)^{1-\mathcal D(h)}}{\left[
1+\left(\frac{\ell}{\eta(h)}\right)^{-2}\right]^{(\mathcal
D(h)-1)/2}}\mbox{ ,}
\end{equation}
where $\mathcal Z (\ell)$ is a normalization factor such that
$\int_{h_{\min}}^{h_{\max}} \mathcal P_\ell(h,\mathcal
R_e/\mathcal R^*,\mathcal D) dh= 1$.

From Eqs. (\ref{eq:betah}) and (\ref{eq:phmeneeul}), one can
derive analytical predictions for the moments of the longitudinal
velocity increment modulus, i.e. $\langle|\delta_\ell
u|^q\rangle$, the energy power spectrum (which is linked to the
Fourier transform of the second order moment
\cite{LohMul95,ChePhD}) and the symmetric part of the longitudinal
velocity increments PDF. This approach has also been successfully
applied in the Lagrangian framework in which the PDFs are
symmetric \cite{CheRou03_1,CheRou03_2}.

\begin{figure*}[t]
\center{\epsfig{file=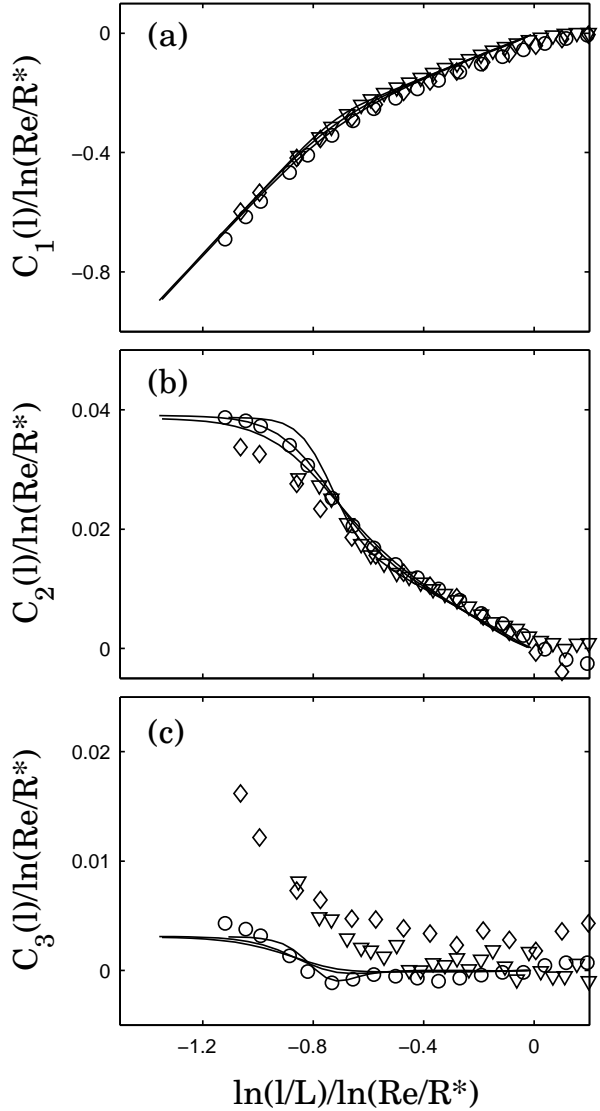,width=8cm}}
\caption{\label{fig:Cumul} Magnitude cumulant analysis of various
experimental longitudinal velocity profiles : ($\diamond$)
Turbulent low temperature gaseous helium jet for $\mathcal
R_\lambda=208$ \cite{ChaCha00}; ($\circ$) Air Jet for $\mathcal
R_\lambda=380$ \cite{baudet}; ($\triangledown$) Wind tunnel for
$\mathcal R_\lambda=2500$ \cite{KahMal98}. (a) $C_1(\ell)=\langle
\ln\beta_\ell\rangle=\langle \ln | \delta_\ell u| \rangle -
\langle \ln |\delta|\rangle$. (b) $C_2(\ell)=\mbox{Var}(
\ln\beta_\ell)=\langle \ln ^2 \beta_\ell \rangle - \langle \ln
\beta_\ell \rangle ^2 =\mbox{Var}( \ln|\delta_\ell u|)-\mbox{Var}(
\ln|\delta|)$. (c) $C_3=\langle \ln^3 \beta_\ell\rangle - 3\langle
\ln^2 \beta_\ell\rangle\langle \ln \beta_\ell\rangle+2\langle \ln
\beta_\ell\rangle^3$. The solid curves correspond to our
theoretical predictions (see text).}
\end{figure*}
As advocated in Ref. \cite{DelMuz}, the magnitude cumulant
analysis provides a more reliable alternative to the structure
function method. The relationship between the moments of
$|\delta_\ell u|$ and the cumulants $C_n(\ell)$ of $\ln
|\delta_\ell u|$ reads
\begin{equation}\label{eq:linkstruccum}
\langle |\delta_\ell u|^p\rangle = \exp \left( \sum_{n=1}^\infty
C_n(\ell)\frac{p^n}{n!} \right) \mbox{ .}
\end{equation}

In Fig. \ref{fig:Cumul}, we report the results of the computation
of the magnitude cumulants $C_n(\ell)$ of various experimental
velocity signals. Actually we have plotted the cumulants of $\ln
\beta_\ell$ instead of $\ln|\delta_\ell u|$, so that they vanish
at the correlation length scale $L$ (as the signature of Gaussian
statistics). When both the cumulants and $\ln (\ell/L)$ are
renormalized by $\ln(\mathcal R_e/\mathcal R^*)$, all the curves
collapse on a universal linear function in the inertial range
(when $\ln (\ell/L)/\ln(\mathcal R_e/\mathcal R^*) \gtrsim -3/4$,
see \cite{CheCas03}) of slope $c_n$. Let us notice that in this
representation, for any Reynolds number, $\ln
(\eta_K/L)/\ln(\mathcal R_e/\mathcal R^*) = -3/4$ and $\ln
(\lambda/L)/\ln(\mathcal R_e/\mathcal R^*) = -1/2$, where $\eta_K$
and $\lambda$ are respectively the Kolmogorov and the Taylor
scales. For the first order cumulant (Fig. \ref{fig:Cumul}(a)),
disregarding large scale anisotropy leading to nonuniversal
effects, $c_1$ is found very close to $1/3$, consistently with K41
theory \cite{Fri95}. For the second-order cumulant (Fig.
\ref{fig:Cumul}(b)), the intermittency coefficient $c_2=0.025\pm
0.003$ is found universal, i.e. independent of the Reynolds number
and of the experimental configuration. For the third one, $c_3$
cannot be claimed to be different from zero (especially at high
Reynolds number) confirming that the statistics of longitudinal
velocity increments are likely to be log-normal
\cite{DelMuz,ArnMan99}. In the intermediate dissipative range
(i.e. $-1.1 \lesssim \ln (\ell/L)/\ln(\mathcal R_e/\mathcal R^*)
\lesssim -3/4$), $C_1(\ell)$ crosses over towards trivial scaling;
the straight line of slope unity observed at smaller scales means
that velocity increments become proportional to the scale (Taylor
development). The behavior of the second-order cumulant is much
more interesting and has been widely studied in Ref.
\cite{CheCas03} : a non trivial Reynolds dependent rapid increase
occurs in the intermediate dissipative range, the larger the
Reynolds number, the more ``rapid" the increase. Finally,
$C_2(\ell)/\ln(\mathcal R_e/\mathcal R^*)$ tends toward a
universal value in the far-dissipative range. Note that
$C_3(\ell)$ displays similar behavior. In Fig. \ref{fig:Cumul} are
also represented our theoretical predictions obtained from the
computation of the moments of $\ln \beta_\ell$ using Eqs.
(\ref{eq:betah}) and (\ref{eq:phmeneeul}), i.e. $\langle (\ln
\beta_\ell)^n\rangle = \int _{h_{\min}}^{h_{\max}} (\ln
\beta_\ell)^n \mathcal P_\ell (h) dh $. We have used the following
set of parameters: $\mathcal R^*=52$ and a universal log-normal
parabolic $\mathcal D (h)$ function,
\begin{equation}\label{eq:DHLN}
\mathcal D (h) =1-\frac{(h-c_1)^2}{2c_2}\mbox{ ,}
\end{equation}
with $c_2=0.025$ and $c_1=1/3+3c_2/2\approx 0.37$ to ensure that
$\zeta_3=3c_1-9c_2/2=1$ in the inviscid limit \cite{Fri95}. The
integration limits $h_{\min}$ and $h_{\max}$ are respectively the
minimal and maximal values such that $\mathcal D(h)\ge 0$. Using
Eq. (\ref{eq:DHLN}), we get $h_{\min} = c_1-\sqrt{2c_2} \approx
0.15$ and $h_{\max} = c_1+\sqrt{2c_2}\approx 0.59$. The different
curves so-obtained superimpose remarkably well with the
corresponding data for the first two cumulants which demonstrates
that our multifractal description accounts quantitatively well for
the departure from scaling in the intermediate dissipative range.
Finite Reynolds number effects \cite{ArnMan99}, statistical
convergence and lognormal approximation can explain some
discrepancies between our theoretical prediction and the behavior
of the third-order cumulant.

\section{Multifractal prediction of the skewness of longitudinal velocity increments}

\begin{figure}[t]
\center{\includegraphics{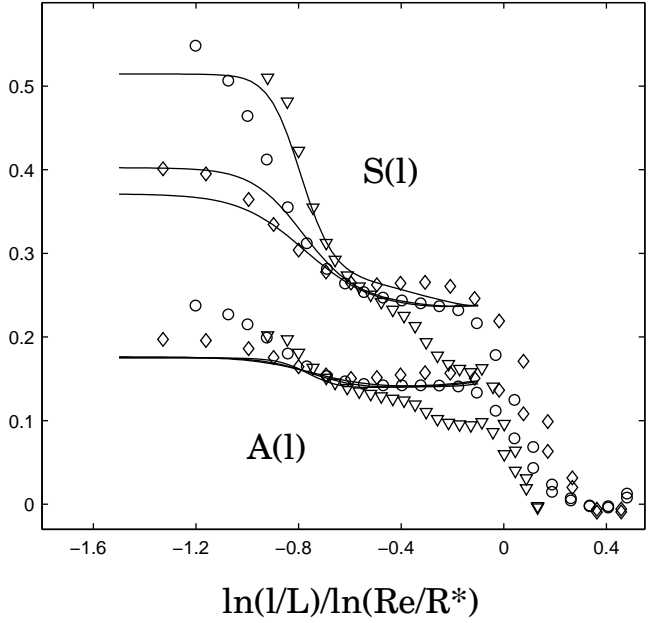}} \caption{Asymmetry factor
$\mathcal A(\ell)$ and Skewness $\mathcal S(\ell)$ estimated from
various experimental velocity profiles. The symbols have the same
meaning as in Fig. \ref{fig:Cumul}. The solid curves correspond to
our theoretical predictions (see text).} \label{fig:Asym}
\end{figure}

Let us now investigate the statistics of signed longitudinal
velocity increments through the two estimators: (i) the skewness
$\mathcal S(\ell) = -\langle (\delta_\ell u)^3\rangle/\langle
(\delta_\ell u)^2\rangle^{3/2}$ and (ii) the asymmetry factor
$\mathcal A(\ell)=-\langle (\delta_\ell u)^3\rangle/\langle
|\delta_\ell u|^3\rangle$. The experimental estimates of $\mathcal
S(\ell)$ and $\mathcal A(\ell)$ are shown in Fig. \ref{fig:Asym}
in a semi-logarithmic representation. Interestingly, $\mathcal
A(\ell)$ displays a plateau at about 0.14 in the inertial range,
whereas the skewness behaves approximatively as a power law. In
the intermediate dissipative range, both estimators undergo a
rapid acceleration, very much like what was observed for
$C_2(\ell)$ in Fig. \ref{fig:Cumul}(b). From a theoretical point
of view, the third-order structure function is solution of the
K\'arm\'an-Howarth-Kolmogorov equation \cite{MonYag71}
\begin{equation}\label{eq:KarHow}
\langle (\delta_\ell u)^3\rangle = -\frac{4}{5}\langle \epsilon
\rangle \ell + 6\nu \frac{d\langle (\delta_\ell
u)^2\rangle}{d\ell}\mbox{ .}
\end{equation}
This equation allows us to compute the third-order structure
function when the second-order one and the average local
dissipation are known. A similar approach has been performed by
Qian in Ref. \cite{Qia00} without taking into account the
intermittency corrections.

We have superimposed in Fig. \ref{fig:Asym} our theoretical
predictions to the experimental data for $\mathcal S(\ell)$ and
$\mathcal A(\ell)$. Indeed, from Eqs. (\ref{eq:betah}) and
(\ref{eq:phmeneeul}), one can compute any moment of the magnitude
of velocity increments $\langle |\delta_\ell u |^n\rangle$ and
velocity gradient $\langle |\partial _xu |^n\rangle$. In
particular, we get
\begin{equation}\label{eq:PredDiss}
\langle \epsilon\rangle = 15\nu \langle (\partial _x u )^2\rangle
= \frac{\sigma^3}{L}\frac{15}{\mathcal R^*}\times \frac{\mathcal
R^* }{\mathcal R_e} \frac{1}{\mathcal Z(0)
}\int_{h_{\min}}^{h_{\max}}\left( \frac{\eta(h)}{L}\right)^{2(h-1)
+1-\mathcal D(h)}dh \mbox{ ,}
\end{equation}
where $\mathcal Z(0)$ is the limit when $\ell \rightarrow 0$ of
the normalization factor $\mathcal Z(\ell)$ appearing in Eq.
(\ref{eq:phmeneeul}). Then, by inserting our description of the
second order structure function (Eq. (\ref{eq:linkstruccum}) for
$p=2$) together with the prediction of the average local
dissipation (Eq. (\ref{eq:PredDiss})) in Eq. (\ref{eq:KarHow}), we
get the third order moment of velocity increments $\langle
(\delta_\ell u )^3\rangle$ at any scale and Reynolds number. As
shown in Fig. \ref{fig:Asym}, the agreement is very good for
distances in between the Kolmogorov and Taylor scales
$\eta_K\le\ell\le \lambda$ (without any arbitrary shifts) when
using the same parameter function $\mathcal D(h)$ and $\mathcal
R^*$ as in our former magnitude cumulant analysis.

Furthermore, we can derive that, when neglecting intermittency
corrections and setting the viscosity to zero in Eq.
(\ref{eq:KarHow}), we get $\mathcal A(\ell) =
3\sqrt{2\pi}/\mathcal R^*\approx 0.145$ and $\mathcal S(\ell) =
12/\mathcal R^* \approx 0.23$, in perfect agreement with
experimental findings. Some discrepancies occur for $\ell\ge
\lambda$, especially for Modane's longitudinal velocity profile,
because (i) of the lack of statistics and (ii) at these scales,
one has to take into account in Eq. (\ref{eq:KarHow}) fluctuations
of the injection rate of energy \cite{LinLun_1,LinLun_2}. In the
intermediate and far dissipative range, our formalism predicts a
universal plateau and a Reynolds number dependence for
respectively the asymmetry factor and the skewness of derivatives,
in consistency with Nelkin's predictions \cite{Nel90} (see Tab.
\ref{tab:tab2} for precise values). We derive in the appendix
\ref{appeSkew} the multifractal prediction for the third order
moment of the velocity gradient $\langle (\partial _x
u)^3\rangle$. Experimentally speaking, measuring gradients is
still controversial mainly because hot wire probe sizes are in
general of the order of the Kolmogorov scale
\cite{ExpSkew_1,ExpSkew_2,ExpSkew_3,ExpSkew_4}. We hope that
further experimental studies will provide definite test of the
validity of our predictions. These preliminary tests are
nevertheless very satisfactory.

\section{Modeling the asymmetric part of the PDFS}

Let us finally elaborate a formalism to describe the PDF of the
signed longitudinal velocity increments. To do so, we suggest to
model the (signed) longitudinal velocity increments in the
following way: $\delta_\ell u = \beta_\ell \times \Delta_\ell$ (in
law), where the positive random variable $\beta_\ell$ is unchanged
but $\Delta_\ell$ is now an independent zero mean random variable
of variance $\sigma^2$, whose probability $\mathcal
P_{\Delta_\ell}$ a priori depends on the scale $\ell$. It follows
that
\begin{equation}\label{eq:CompPDF}
\mathcal P_{\delta_\ell u}(\delta_\ell u)  =
\int_{h_{\min}}^{h_{\max}} \frac{dh}{\beta_\ell (h)}\mathcal
P_\ell (h) \mathcal P_{\Delta_\ell} \left( \frac{\delta_\ell
u}{\beta_\ell}\right) \mbox{ .}
\end{equation}
According to the Edgeworth's development \cite{McC87}, any PDF can
be decomposed over a ``basis" of the successive derivatives of a
Gaussian function:
\begin{equation}\label{eq:Edge}
\mathcal P_{\Delta_\ell}(x) = \sum_{n=0}^{+\infty}\lambda_n(\ell)
\frac{d^n}{dx^n}\left(
\frac{1}{\sqrt{2\pi\sigma^2}}e^{-x^2/2\sigma^2}\right) \mbox{ .}
\end{equation}
\begin{figure}[t]
\centerline{\includegraphics[width=9cm]{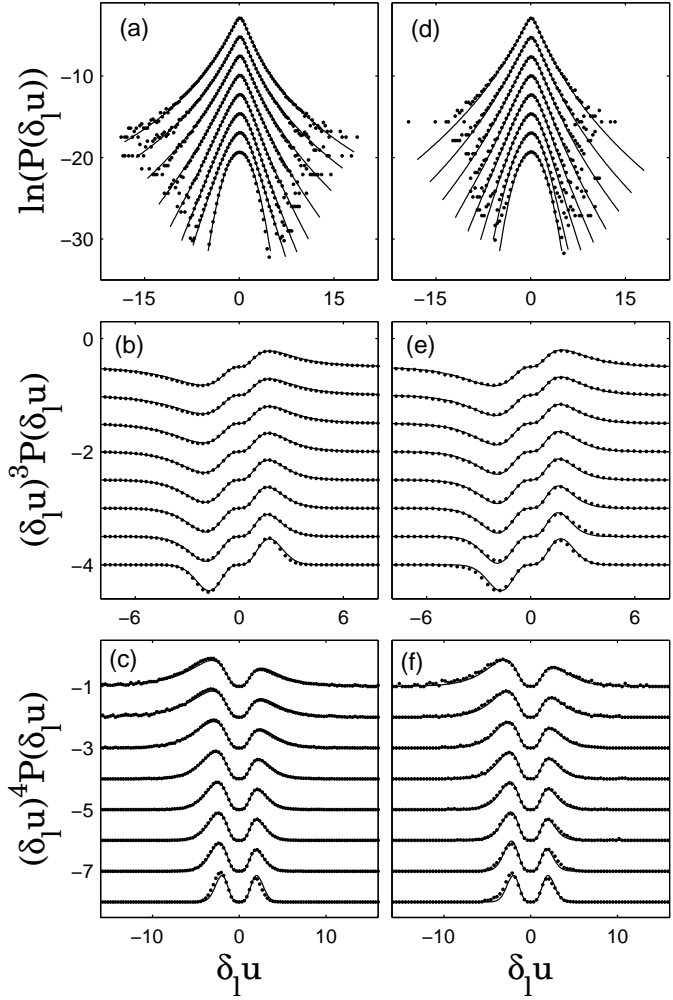}}
\caption{\label{fig:PDFs} PDFs of signed longitudinal velocity
increments of Air Jet \cite{baudet} (a-c) and Modane
\cite{KahMal98} (d-f) velocity signals for various scales. (a) and
(d) : $\ln \mathcal P_\ell (\delta_\ell u)$. (b) and (e) : $
(\delta_\ell u)^3\mathcal P_\ell (\delta_\ell u)$. (c) and (f) : $
(\delta_\ell u)^4\mathcal P_\ell (\delta_\ell u)$. Represented
scales (from top to bottom) :  $\ln(\ell/L)=$
-6.0069,-5.3137,-4.6206,-3.9274,-3.2343,-2.5411,-1.8480,0.9246 for
the Air Jet data and $\ln(\ell/L) =$ -6.4137, -5.6028, -4.6645,
-3.6411, -2.7501, -1.8598, -0.8685, 0.1226 for Modane data. All
curves are arbitrarily vertically shifted for the sake of clarity.
The solid curves correspond to our theoretical predictions (see
text).}
\end{figure}
The symmetric part (even terms) of the PDF of $\Delta_\ell$, i.e.
$\mathcal P^+_{\Delta_\ell}$, is well described by a Gaussian
$\delta$ noise (as previously stated), which means that $\lambda_0
(\ell) = 1$ and $\lambda_{2n}(\ell)=0$ for $n\ge 1$ and every
scale $\ell$. Furthermore, it can be demonstrated that whatever
$\sigma^2$ is, $\langle(\delta_\ell u)^3\rangle = -6
\lambda_3(\ell)\langle (\beta_\ell)^3\rangle$. Hence,
$\lambda_3(\ell)$ is fully determined by the
K\'arm\'an-Howarth-Kolmogorov equation (Eq. (\ref{eq:KarHow})). As
Eq. (\ref{eq:KarHow}) is the only available constraint on
$\lambda_n$, it is quite natural (as a first approximation) to
restrict the expansion to $\lambda_3$: $\lambda_{2n+1}(\ell)=0$
for $n\ge 2$ and every scale $\ell$. Additional statistical
equations involving odd moments of $\delta_\ell u$ would be needed
to give the next $\lambda_{2n+1}(\ell)$. This would require
further modeling (primarily to get ride of pressure terms), which
is out of the scope of the present work. Unfortunately, the
previous crude approximation for the odd terms of (\ref{eq:Edge})
leads to severe pathologies, such as negative probability for rare
large $\Delta_\ell$ events and is not consistent with higher-order
statistics as Hyperskewness $\langle (\delta_\ell u)^5\rangle/
\langle (\delta_\ell u)^2\rangle ^{5/2}$ (data not shown).
Nevertheless, since the third-order structure function does not
depend on the precise variance $\sigma^2$ entering in the third
order derivative of the Gaussian PDF, we propose to renormalize
the variance $\sigma^2$ of the retained odd term ($n=3$): $\tilde
\sigma^2 = 0.9 \sigma^2$. We thus obtain $\mathcal
P_{\Delta_\ell}(\Delta_\ell) = \mathcal
P^+_{\Delta_\ell}(\Delta_\ell) + \mathcal
P^-_{\Delta_\ell}(\Delta_\ell)$, where the asymmetric part of the
PDF of $\Delta_\ell$ is modeled as
\begin{equation}
P^-_{\Delta_\ell}(x) = \lambda_3(\ell)
\frac{d^3}{dx^3}\left(\frac{1}{\sqrt{2\pi\tilde\sigma^2}}e^{-x^2/2\tilde
\sigma^2} \right)\mbox{ .}
\end{equation}

The PDFs of longitudinal velocity increments so-obtained from Eq.
(\ref{eq:CompPDF}) are shown in Fig. \ref{fig:PDFs} for several
scales spanning the inertial and intermediate dissipative ranges,
and compared to the experimental ones for both Air jet Fig.
\ref{fig:PDFs}(a-c) and Modane \ref{fig:PDFs}(d-f). We see a
continuous deformation across scales, from Gaussian at the
correlation length scale ($L$) to exponential-like distributions
in the inertial range and ultimately to stretched-exponential when
dissipation starts acting. This PDF shape evolution is the
signature of intermittency and is remarkably reproduced by our
formalism when using the same quadratic parameter $\mathcal D(h)$
function and constant $\mathcal R^*=52$ as in Figs.
\ref{fig:Cumul} and \ref{fig:Asym}. This agreement is emphasized
in Figs. (\ref{fig:PDFs})(b,d) and (\ref{fig:PDFs})(c,f) where
respectively $ (\delta_\ell u)^3\mathcal P_\ell (\delta_\ell u)$
and $ (\delta_\ell u)^4\mathcal P_\ell (\delta_\ell u)$ are
represented as a quantitative test of the relevance of our
formalism to account for the dissymmetrical PDF tails.

\section{Conclusion}

To conclude, we have shown that the evolution across scales of the
signed longitudinal velocity increments statistics, from the
inertial far down the dissipative ranges, depends only on a
universal parameter function $\mathcal D(h)$ and a universal
constant $\mathcal R^*$ that must be seen as a multifractal
version of the Kolmogorov constant. In particular, neglecting the
intermittency corrections, we provide an enhanced phenomenology of
turbulence in deriving the value of the Skewness $\mathcal S(\ell)
= 12/\mathcal R^* \approx 0.23$ in the inertial range. We have
further shown that choosing a quadratic form for $\mathcal D(h)$
(i.e. the hallmark of an underlying lognormal cascading process)
provides a very good quantitative description of the longitudinal
velocity increments PDFs measured in several flows, in different
geometries and for different Reynolds numbers. This study proposes
a new formalism, relying on the Edgeworth's development, which
opens the route to the modeling of velocity increment PDF. New
experimental investigations of velocity gradients statistics would
be welcomed as an additional and complementary test of our
theoretical multifractal approach.

We wish to acknowledge P. Flandrin for fruitful discussions.

\appendix

\section{Multifractal prediction of the Skewness of derivatives}

In an infinite domain, or in a finite domain with periodic
boundary conditions, a Taylor's development of the second order
structure function leads to
\begin{equation}\label{eq:TayDevStruc2}
\langle (\delta_\ell u)^2\rangle = \langle \left(
\sum_{n=1}^4\partial^n_xu\frac{\ell^n}{n!}+o(\ell^4)\right)^2\rangle
= \ell^2\langle (\partial_xu)^2\rangle -\frac{\ell^4}{12}\langle
(\partial^2_xu)^2\rangle +o(\ell^4)\mbox{ .}
\end{equation}
Inserting the development pointed by Eq. (\ref{eq:TayDevStruc2})
into the K\'arm\'an-Howarth-Kolmogorov equation (Eq.
(\ref{eq:KarHow})) leads to
\begin{equation}\label{eq:secondorderder}
\langle (\partial_xu)^3\rangle=-2\nu \langle
(\partial^2_xu)^2\rangle \mbox{ .}
\end{equation}
This classical result can be found in Ref. \cite{MonYag71}. One
may wonder whether our description of the second order structure
function (Eqs. (\ref{eq:betah}) and (\ref{eq:phmeneeul})) is
consistent with this development (Eq. (\ref{eq:TayDevStruc2})). In
particular, the pre-supposed continuous and differentiable
transition between the inertial and the dissipative range of scale
inspired from the work of Batchelor \cite{Bat51} should give a
leading term proportional to $\ell ^3$ once inserted in Eq.
(\ref{eq:KarHow}). This property constrains seriously the possible
form of the transition. The transition form used here benefits of
such property. We get, with the help of a symbolic calculation
software,
\begin{equation}\label{eq:pred1}
\langle (\partial_xu)^3\rangle = -\frac{6\nu\sigma^2}{L^4} \left[
\frac{2}{\mathcal Z(0)}
\int_{h_{\min}}^{h_{\max}}\left[2h-1-\mathcal D(h) \right]\left(
\frac{\eta(h)}{L}\right)^{2(h-2)+1-\mathcal D(h)}dh + \mathcal F
\right]\mbox{ ,}
\end{equation}
where $\mathcal F$ is a negligible additive term, coming from the
Taylor's development of the normalization factor $\mathcal
Z(\ell)$, and given by
\begin{equation}
\mathcal F = -\frac{1}{\mathcal Z(0)
^2}\int_{h_{\min}}^{h_{\max}}\left[1-\mathcal D(h) \right]\left(
\frac{\eta(h)}{L}\right)^{-1-\mathcal D(h)}dh
\int_{h_{\min}}^{h_{\max}}\left(
\frac{\eta(h)}{L}\right)^{2h-1-\mathcal D(h)}dh \mbox{ .}
\end{equation}
\begin{table}
\begin{center}
\begin{tabular}{lcr}
$\mathcal R_\lambda$&$\mathcal S(0)$&$\mathcal A(0)$\\
\hline
208 &  0.35 & 0.17\\
380 & 0.38 & 0.17\\
2500 & 0.50 & 0.17\\
\end{tabular}
\caption{\label{tab:tab2}Theoretical predictions of the Skewnesses
and Asymmetric factors of velocity derivatives (Eq.
(\ref{eq:pred1})) for the three different experiment longitudinal
velocity data sets presented in Figs. \ref{fig:Cumul} and
\ref{fig:Asym}. We have used $\mathcal R^*=52$ and a parabolic
function for $\mathcal D(h)$ (Eq. (\ref{eq:DHLN})) with parameters
$c_2=0.025$ and $c_1=1/3+3c_2/2$.}
\end{center}
\end{table}

Eq. (\ref{eq:pred1}) can be seen as the multifractal prediction of
the third order moment of the velocity derivatives, using the same
transition interpolation form as in Eqs. (\ref{eq:betah}) and
(\ref{eq:phmeneeul}). This is also a prediction for the second
order moment of the second order derivative of velocity via Eq.
(\ref{eq:secondorderder}). We gather in Table \ref{tab:tab2} the
theoretical values for the Skewness $\mathcal S(0) = \langle
(\partial_xu)^3\rangle/\langle (\partial_xu)^2\rangle^{3/2}$ and
the Asymmetric factor $\mathcal A(0) = \langle
(\partial_xu)^3\rangle/\langle (\partial_xu)^2\rangle^{3/2}$, i.e.
the limit when $\ell \rightarrow 0$ of our theoretical predictions
for the Skewness and Asymmetric factor of velocity increments
presented in Fig. \ref{fig:Asym}. A specifically designed
experiment aimed at measuring the fluctuations of longitudinal
velocity increments for scale much smaller than the Kolmogorov
length scale will provide a decisive test of the validity of these
theoretical predictions.

\label{appeSkew}

\end{document}